\documentclass[usenatbib,twocolumn]{mn2e}
\usepackage{graphicx}
\usepackage{color}
\usepackage{amssymb}
\usepackage{subfigure}
\usepackage{wasysym}
\newcommand{\beq}{\begin{equation}}
\newcommand{\eeq}{\end{equation}}

\def\beq{\begin{equation}}
\def\ee{\end{equation}}
\def\lsim{\mathrel{\rlap{\lower4pt\hbox{\hskip1pt$\sim$}}
    \raise1pt\hbox{$<$}}}
\def\gsim{\mathrel{\rlap{\lower4pt\hbox{\hskip1pt$\sim$}}
    \raise1pt\hbox{$>$}}}

 \title[Effect of planetary magnetic fields on wind capture]
 {Mass, energy, and momentum capture from stellar winds by magnetized and unmagnetized planets: implications for atmospheric erosion and habitability}

\author [Blackman,Tarduno]
{Eric G. Blackman$^{1,2,3}$\thanks{E-mail: blackman@pas.rochester.edu},
John A. Tarduno$^{1,3}$\thanks{E-mail: john.tarduno@rochester.edu} 
\\ $^{1}$Department of Physics and Astronomy, University of Rochester, Rochester NY, 14627, USA\\
 $^{2}$Laboratory for Laser Energetics,  University of Rochester, Rochester NY, 14623, USA\\
 $^{3}$Department of Earth and Environmental Sciences, University of Rochester, Rochester NY, 14627, USA\\ }
\begin{document}
\date{}
\pagerange{\pageref{firstpage}--\pageref{lastpage}} \pubyear{}
\maketitle
\label{firstpage}


 \begin{abstract}
The extent to which a magnetosphere protects its planetary atmosphere  from  stellar wind ablation depends  upon how  well it prevents energy and momentum exchange with the atmosphere and how well it traps otherwise escaping plasma. We focus on the former, and provide a formalism for  estimating  approximate upper limits on mass,  energy, and momentum capture, and use them to constrain  loss rates. Our approach quantifies a  competition between the local deflection of incoming plasma by a planetary magnetic field
and the increase in area for solar wind mass capture provided by this magnetosphere
 when the wind and planetary fields incur magnetic reconnection.  Solar wind capture  can be larger for a magnetized planet versus an unmagnetized planet with small ionosphere, even if the rate of  energy transfer is  less.  We find this to be the case throughout Earth's  history since the lunar forming impact-- the likely start of the geodynamo.  Focusing of incoming charged particles near the magnetic poles can  increase the local energy flux for a strong enough planetary field and weak enough wind, and we find  that this may be the case for  Earth's future. The  future protective role of Earth's  magnetic field would then depend  on its ability to trap outgoing plasma. This competition between  increased collection area and reduced inflow speed from a magnetosphere  is likely  essential in determining the net protective role of planetary   fields and its importance in sustaining habitability, and can help explain the current ion loss rates from Mars, Venus and Earth. 
\end{abstract}

\begin{keywords} 
planets and satellites: magnetic fields;
planets and satellites: terrestrial planets; 	 
stars: winds, outflows 	
Earth; 
stars: activity; 
planet-star interactons
\end{keywords}

\section{Introduction}

Traditionally,  planetary habitable zones are defined as the range of distances from a star within which a planet can orbit such  that  stellar luminosity heats the planet to a temperature range that can sustain gravitationally bound liquid water \citep{Huang1960,Kasting1993}.  However, even if  liquid water is essential,  the potentially catastrophic effect of stellar activity  from ultraviolet and X-ray photons, and   stellar winds on the planetary atmosphere must also be
considered \citep[e.g.][]{Lammer2013}. High energy photons can  photo-dissociate molecules and/or  heat atoms such that  that the atmospheric constituents exceed the escape speed at the exobase  and are catastrophically lost.
 Complementarily, a  strong stellar wind can exchange momentum with the planetary atmosphere and directly or indirectly eject the essential constituents. 

 The overall effects of stellar activity on atmospheres are commonly 
(if not misleadingly) delineated as ``thermal" or ``non-thermal" mechanisms 
 \citep[e.g.][]{Hunten1982,Kumar1983+,Kasting1983+,Shizgal1996+,Lammer1991+,Hunten1993,Seager2010,Lammer2013}.  For the former,  energy is primarily absorbed and converted to thermal energy.
 The potential for ejection amounts to a comparison between thermal velocities and escape speed.
Non-thermal mechanisms refer to those in which the input energy is not thermalized by collisions and/or may involve a change of chemistry  or ionization  state of the influenced particles.   Erosion of atmospheres by  stellar winds is an important example of the latter,
and the focus of the present paper.  Ultimately, a  realistic determination of habitability will depend  on understanding the high energy coronal activity from stars and the efficacy with which  planetary atmospheres are protected against the associated radiation and winds.

 The magnetic field strength and X-ray activity of solar-like stars 
increase with rotation rate for slow rotators and
then saturate for very fast rotators \citep{Noyes1984,Schrijver2000,Pizzolato2003,Mamajek2008,Wright2011,Vidotto2014, Reiners2014}.
 Relationships between stellar activity, rotation, and magnetic field strength are all connected   to the dynamo origin of these fields within the stars; it is an ongoing enterprise  of research to understand these connections and make them quantitative \citep{Karak2014,Kitchatinov2015,Sood2016,Blackman2016}.
In the absence of significant cooling, the winds from solar-like stars, which also draw from the coronal magnetic and thermal energy, are also likely significantly stronger for younger stars.
 This too, however, is debated and a topic of active research  \citep{Wood2006,Wood2014,Cohen2011,Suzuki2013,Matt2015,Blackman2016}.

Planetary magnetospheres produced from core dynamos are widely perceived to be protective against erosion from stellar winds. 
This is of topical interest in understanding the evolution of the early Earth \citep{Tarduno2015+} and well as the potential habitability of exoplanets \citep{Nascimento2016+}.  The  planetary magnetic field  acts as a magnetic cage that tempers stripping by stellar winds, redirecting  flow  starting at a bow shock, and  diverting it around  the planet. The bow shock lies just outside the radius of the ``obstacle" namely the magnetopause, where the ram pressure of the stellar wind is balanced by the magnetospheric pressure of the planetary atmosphere. For example, the difference between the current water inventory of Earth versus that of  Mars has long been  thought to be due to the presence of an internally generated magnetic field on the   former, and the loss of water on the latter \citep[e.g.][]{Luhmann1992+,Lundin2007}.


However, some observations have also led to  speculation that the  magnetosphere might not be exclusively protective  \citep{Brain2013}.  
For example,   
the present-day total atmospheric ion escape from Earth, Mars and Venus appears to be within similar orders of magnitude, at 10$^{24}$-10$^{26}$ ions s$^{-1}$ 
\citep{Barabash2010, Strangeway2010, Dubinin2011+,Wei2012,Jakosky2015+,Strangeway2017+},  but here too there is debate fueled by the limitations of observations and models.  Because Earth has an internal geomagnetic field and Mars and Venus do not,  the role of the internal magnetic field might seem to be  un-influential.
 Note that understanding the  present kinematic loss rates 
and the  role of the magnetic field does not by itself 
determine  relative habitability. The latter also depends on the long term
history of each planetary atmosphere  that includes its chemistry  and in particular, water loss \citep{Tarduno2014}, which is crucial for life.

The competing  influences of planetary magnetic fields are also manifest in their role of channeling ions. 
Incoming  ions are channeled toward the poles in a large scale dipole configuration. In this respect,  magnetic fields might exacerbate atmospheric loss rather than abate it \citep{Brain2013}.   Such an effect would preferentially
eject heavy ions that would otherwise be gravitationally bound at atmospheric temperatures, unlike light ions like 
H$^{+}$ whose thermal speed would  already exceed  the escape speed.
Supporting evidence comes from
observations  \citep{Strangeway2005}
  showing an enhanced  outward  oxygen ion flux from auroral zones  consistent with 
  a two order-of-magnitude concentration of  solar wind energy     \citep{Moore2010}.
On the other hand, dipole magnetic fields 
can also recapture at least some of this outgoing flux. For example, \cite{Seki2001} conclude that the polar outflow of O$^{+}$ from Earth today is some 9 times greater than the net loss, considering recapture in the magnetosphere.  A local outflow might therefore not always result in a catastrophic global outflow.  Complementarily, the depth to which incoming ions can penetrate also depends on competing forces such  as a magnetic mirror force
 \citep{Cowley1990+}. 
 
Despite  the complexities that determine the actual fate and influence of the incoming particle flux, the associated gain in energy and momentum of the  atmosphere cannot
exceed that supplied by the  wind. 
This in turn leads to  an approximate upper limit on the potential atmospheric mass  loss (as we shall see) if the energy or momentum conversion into escaping particles were maximally efficient and the atmospheric  particles were not already hovering close to the escape speed before the influence of the wind.  It is both instructive, and important even, to constrain how an internally produced magnetic field affects this approximate upper limit compared to a planet without an internal magnetic field. As we will see, even this simple comparison provides some conceptual and quantitative insight.

In section 2 we estimate   wind  mass and energy  supply rates entering  planetary atmospheres for ``magnetized" vs. ``unmagnetized" planets (with these terms to be defined more carefully therein).  We  take into account magnetic reconnection when the stellar wind field is anti-aligned with the planetary field and quantify the competing effects of reduced inflow speed and increased collection area. We also include the effects of polar focusing. 
In section 3, we use  the results of section 2   to derive approximate upper limits on the mass loss rates for magnetized vs. unmagnetized planets,  and include estimates for both energy and momentum-limited loss rates.
In section 4 we  apply the results to Earth, Mars, and Venus. Using time-evolution models for the host stellar wind ram pressure to calculate the time-dependent magnetospheric stand-off distance, we show how  the solar wind mass  capture rates throughout Earth's history are actually larger  than those for an unmagnetized Earth-like planet with a small ionosophere, even though total maximum energy capture rates, and the associated  atmospheric mass loss rates would be lower.
This conclusion changes for Earth's future when polar focusing is included.
By  consideration of  momentum transfer from wind to the  atmosphere-- and
specifically to pick-up ions for Venus and Mars-- we predict $O^+$ loss rates of Mars, Venus and Earth  from
our formalism. We  conclude in section 5.


 




\section{ Mass and Energy Loading With and Without A Magnetosphere}
\label{sec2}

\subsection{Standoff distance for planet with dipole field}

For a  planet with a dipole  magnetic field and a stellar wind dominated by  ram pressure,  the standoff distance $d_s=r_s$, where the $r_s$ is determined by
where the magnetic pressure of the planet equals the ram pressure of the impinging stellar wind, and 
is given by 
\beq
{r_s^3\over r_p^3} = \left({B_p^2 \over 8\pi\rho_w v_w^2}\right)^{1/2}, 
\label{4.4.5}
\eeq 
where $r_p$ is the radius of the planet, $B_p$ is the magnetic field strength at the planetary surface, $\rho_w$ and $v_w$ are the stellar wind density and speed at $r_s$ respectively.


\subsection{Incoming mass flux for intrinsically ``unmagnetized" and ``magnetized" planets}
\begin{figure}
\centering
{
\includegraphics[width=1\columnwidth]
{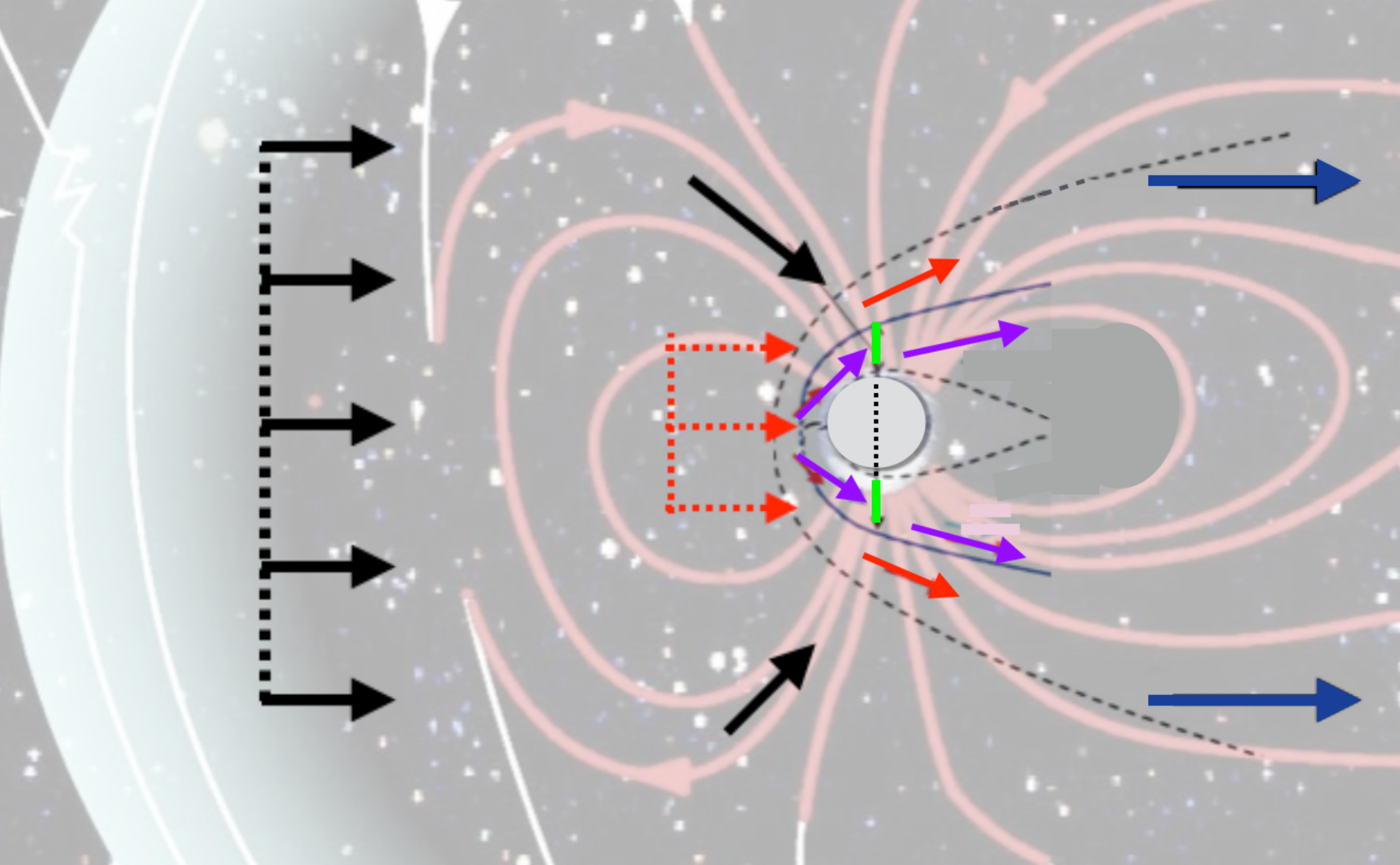}}
\caption{Schematic of both ``magnetized" and ``unmagnetized" cases superimposed with the same sized planet for comparison.
[Adapted into a composite from Dubinin et al (2011) and public domain graphic by A. Krasse (NASA Goddard)]. Stellar wind is incoming from the left.
{\it For the "magnetized" planet case, focus on  the thick black, pink, and blue lines and arrows and ignore the thin lines:} the thick pink lines are the magnetic field of the planet;  the thick black arrows indicate incoming solar wind flow; the thick  blue arrows indicate the outgoing flow from the atmosphere, so colored to indicate that they may also include  solar wind material.  
The projected cross section of the magnetopause/interaction ($\sim \pi r_s^2$, where $r_s$ is defined in  Eq. \ref{4.4.5})  is indicated by the thick vertical dotted black line at the left. The outer bubble is the bow shock and the white lines are the interplanetary field. 
{\it  For the "unmagnetized" case, focus on the thin lines and arrows and ignore the aforementioned thick lines:}   the thin curved dotted line is the bow shock;  the inner dotted line is the boundary of the induced magnetopause and the thin solid line between those two bounds the the region 
 where pickup ions interact with the incoming flow. 
 The red arrows indicate the incoming flow and the violet arrows indicate
 the pickup ion flow swept by the stellar wind.
 The  green distance between the ionopause and  the bow shock can  be of order a gyro-radius away of thermal ions, and increases away  from the dayside ionopause  as the flux pile up decreases and the field weakens to its ambient value.
 The projected cross section  of the magnetopause/interaction area for this "unmagnetized" case ($\sim \pi r_T^2$, see Eq. \ref{rT}),
  is indicated by the dotted vertical red line.}
 \label{schematic}
\end{figure}

The rate of   stellar wind ions entering a  planet's atmosphere depends on the product of the effective area onto which ions are captured and
the speed of the inflow.
We focus  on comparing planets of two extremes which we delineate   as ``magnetized" and ``unmagnetized". The difference is shown schematically in Figure \ref{schematic}, 
where the two  cases  are superimposed. 
For present purposes, ``magnetized" planets  are those
for which the standoff distance is determined by a  magnetopause of radius $r_s >> r_p$.
The thick lines of Fig. \ref{schematic} correspond to this case;  the incoming flow is represented by the black lines and the outgoing mixture of  wind and atmosphere  flow is given by thick blue lines. The cross section for interaction is given by the dotted black line.
  an 
Complementarily,  our ``unmagnetized" case  refers to planets whose standoff  
distance $d_s$ from the incoming wind ram pressure equals the radius $r_i$ set by  ionospheric plus induced magnetic pressure.  The  cross section of interaction  between the stellar wind   and the atmosphere is, however, determined by an interaction radius $r_T\ge r_i$
which  we write as 
 \beq
  r_T=r_i+ r_g
  \label{rT}
  \eeq
   where  $r_g$  is the gyro-radius of the 
  atmospheric  ions, an upper limit of which is that for 
  the uncompressed wind magnetic field at the planet.
  The thin lines of Fig. \ref{schematic} capture this schematically.
  The incoming wind flow arrows are red and the mix of wind and outgoing atmospheric ion flow arrows are purple.
  The green represents  $r_g$ at the terminator such that the cross section of interaction is given  by the red dotted line.   For  Mars and Venus, the ionopause and magnetopause (i.e., the induced magnetopause boundary) are just  a few hundred km above the surface,  and not dramatically delocalized from the exobase radius \citep{Russell2016+}. 
  For oxygen ions, $r_g$ can be
  comparable to, or even significantly larger than the planet radius (particularly for Mars) as we   discuss  further in section 4.2.2.
For a more compact case, it is possible that  $(r_i+r_g)-r_p < r_p$ such that 
  $r_T \sim r_p$.  This does not preclude the planet from still having  a non-trivial atmosphere.
  
For our ``unmagnetized" case then, the incoming mass per unit time potentially influencing  the cross section determined by $r_T$ is given by
\begin{equation}
{\dot M}_{c}\simeq \frac{\pi r_T^2}{4\pi R^2}{{\dot M}_w},
\label{4.4.1}
\end{equation}
where ${\dot M}_w$ is the star's mass loss rate per unit time of the wind,
 and  $R$ is the planet-star distance.

 In contrast, for the ``magnetized" case, we consider particle capture by the magnetosphere to occur primarily during reconnection events between the stellar wind is fand the  planetary magnetosphere. This is because  we are considering  the magnetized case to have a magnetopause so far from the exobase  that the atmospheric density is 
too low for interacting in that region to cause significant atmospheric mass loss. The reconnection
allows the wind plasma closer access to the atmosphere.  Other pathways to particle capture such as Kelvin-Helmholtz mixing or impulsive penetration  can also occur \citep{Gunell2012}, but we  focus on magnetic reconnection as the primary facilitator.

 \cite{Paschmann2013} point out that about 50\% of the time
reconnection is favorable at the Earth's magnetopause.  This is not  directly attributable  to the fact that  after averaging over  many rotations and  cycle periods over which the field periodically reverses,  the shear angle between the wind field and magnetospheric field would be  favorable  to reconnection
approximately half of the time. Rather, the favorability depends on a  combination of shear angle, current sheet thickness, and variation  in the plasma $\beta$ (ratio of plasma to magnetic pressure) across the magnetopause.  Reconnection is still possible with small shear angles (and thus strong guide field) but requires more symmetry in $\beta$ across the magnetopause interface  than the large shear angle cases.    In this respect,
there need  not be a strict correlation between mass loading and field orientation of the solar wind.

During  the  reconnection phase, the effective mass capture area for stellar wind ions can approach the  magnetopause cross section (on the dayside).
Determining the overall efficacy of the magnetosphere at keeping incident wind particles out then requires accounting for the competing effects of the increased  collection area with the reduced speed.  With this in mind, the mass  collected  per unit time ${\dot M}_{c,m}$ is given by 
\begin{equation}
{\dot M}_{c,m}= \frac{\pi r_{s}^2 }{4\pi R^2 } \frac{v_{in}} {v_w}{{\dot M}_w\over 2} 
\label{4.4.2},
\end{equation}
where
$v_{in}$ is the effective inflow speed, taken to be  the reconnection speed $v_{rec}$ derived below.
The factor of  ${{\dot M}_w\over 2} $ 
 accounts for the fact that the incoming wind loads into the atmosphere only when there is reconnection.

The ratio of the mass per unit time captured by a planet with a magnetosphere to that captured by the  same planet without a magnetosphere is given by the ratio of Eq. (\ref{4.4.2}) to Eq.  (\ref{4.4.1}). This is
\begin{equation}
Q\equiv{{\dot M}_{c,m}\over {\dot M}_{c}}= {1\over 2} \left(r_s\over r_T \right)^2 \left(v_{in}\over v_w\right).
\label{4.4.3}
\end{equation}
Not all of the variables in Eq. (\ref{4.4.3}) are independent, but we can simplify crudely as follows:
reconnection  at the standoff point  is generally   asymmetric  across the current sheet and
the  incoming speed of reconnection 
from the wind-side toward the  sheet is given by \citep{Cassak2007+}
\beq
\begin{array}{r}
v_{in}=v_{rec}\sim2 { \tilde \chi}\left({B_{w}B_2^{3} \over4\pi (B_w+B_2) (B_w\rho_2+B_2\rho_w) } \right)^{1/2},
\end{array}
\label{4.4.4b}
\eeq
where $B_w$ is \textbf{the} field on the wind-side of the magnetopause, $B_2$ is the planet field on the
magnetospheric side of the reconnection region, and ${ \tilde\chi}$ is the ratio of current sheet thickness to length. 
 For Earth at least,   data suggest  $B_2 >B_w$ and $B_2\rho_w> B_1\rho_2$ \citep{Sonnerup2016} so
$v_{in} $  reduces to
\beq
\begin{array}{r}
v_{in}=v_{rec}\sim {2 B_{w}^{1/2}B_2^{1/2} \over (4\pi\rho_w)^{1/2} }{\tilde \chi}
 \simeq \chi
\left({B_p \over \sqrt {4\pi \rho_w }}\right)
\left({r_p \over r_s}\right)^3,
\end{array}
\label{4.4.4}
\eeq
where $\chi=2{\tilde \chi}\sqrt{ B_w/B_2}$, and we use the dipole 
scaling $B_2 = B_p ({r_p / r_s})^3$ where $B_p$ is the surface field strength.
 Observations and simulations are roughly consistent  with $\tilde \chi =0.1$   \citep{Cassak2007+}  
 and in turn $0.5\ge \chi\ge 0.1$.

We now combine Eqs.  (\ref{4.4.4}) and (\ref{4.4.5})  with Eq. (\ref{4.4.3})
for the specific case in which $r_T \sim r_p$, the unmagnetized case with a compact atmosphere.
The result is
\beq
v_{in}/v_w ={ \chi} {\sqrt 2},
\label{vrecvin}
\eeq
and
\begin{equation}
Q={\chi\over \sqrt 2} (r_s^2/r_p^2)
=7.07
\left ({\chi\over 0.1}\right)  \left({r_s/ r_p \over 10}\right)^2, 
\label{4.4.5a}
\end{equation}
where we have scaled to present day Earth-like values of $\chi$ and $r_s/r_p$.
That $Q$  can exceed  unity  highlights the important effect  of the increased collecting area that  the magnetosphere provides. 
 Eq. (\ref{4.4.5a}) shows that the weaker the wind, the higher the rate of mass capture for a magnetized planet relative to an  unmagnetized planet since $r_s/r_p$  increases for weaker stellar winds.  In what follows, we assume  $\chi$ to be relatively constant over the range of $r_s/r_p$
 calculated. This is consistent with our assumption  that 
 the current sheet geometry is  relatively constant and the fact that 
 $\xi$ depends on the ratio of   $B_w$ to $B_2$.  Both of these two field strengths are likely larger for younger stars; $B_w$ is correlated with the surface field which is stronger for younger stars with higher coronal thermal power for the wind.  The wind's thermal
 energy    supplies its ram pressure which, in turn, reduces $r_s$ and increases $B_2.$.

 
 
 \begin{figure}
\centering
\vskip-0.2in
{\includegraphics[angle=-90,width=1\columnwidth]
{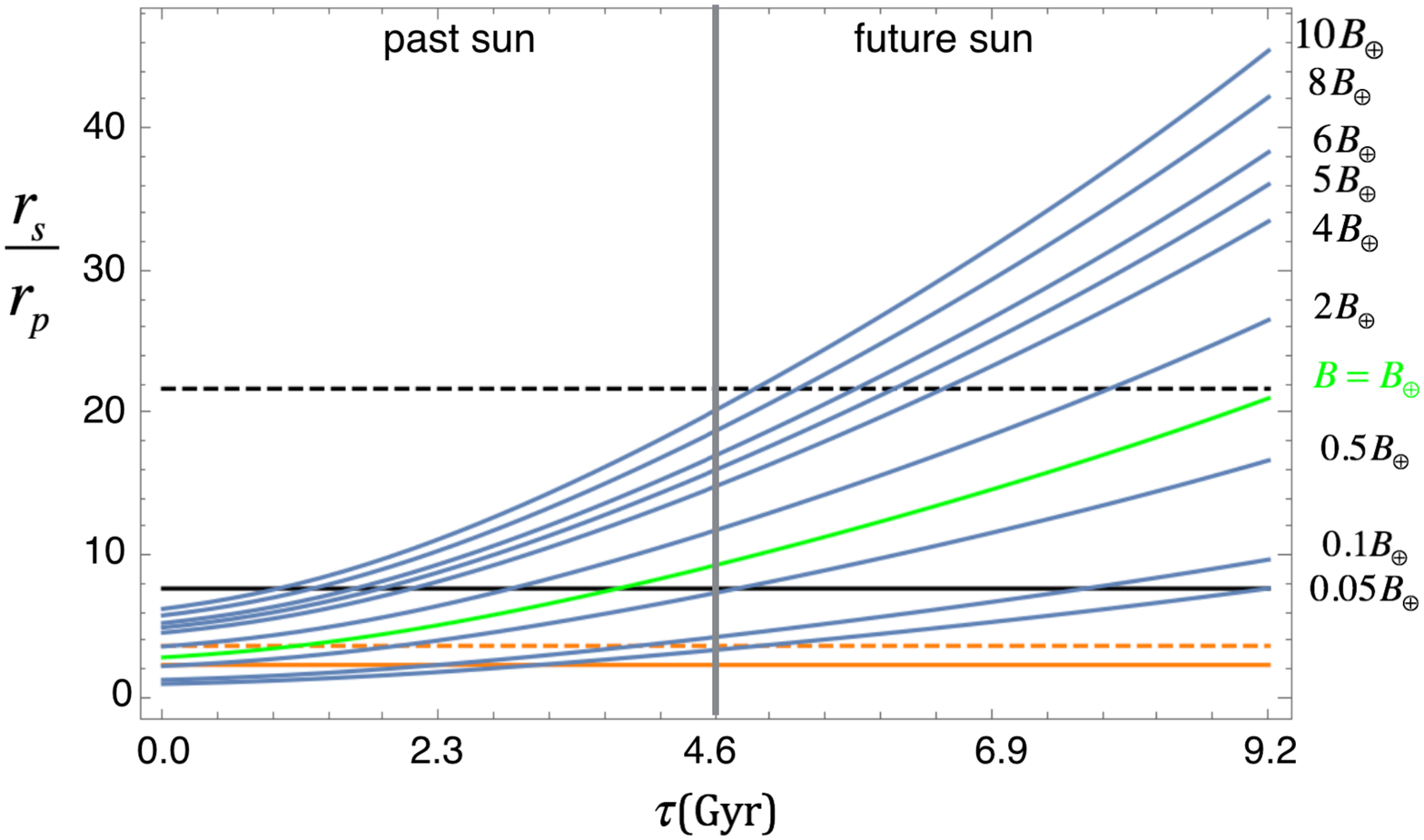}
\vskip-0.4in
\includegraphics[width=0.9\columnwidth]
{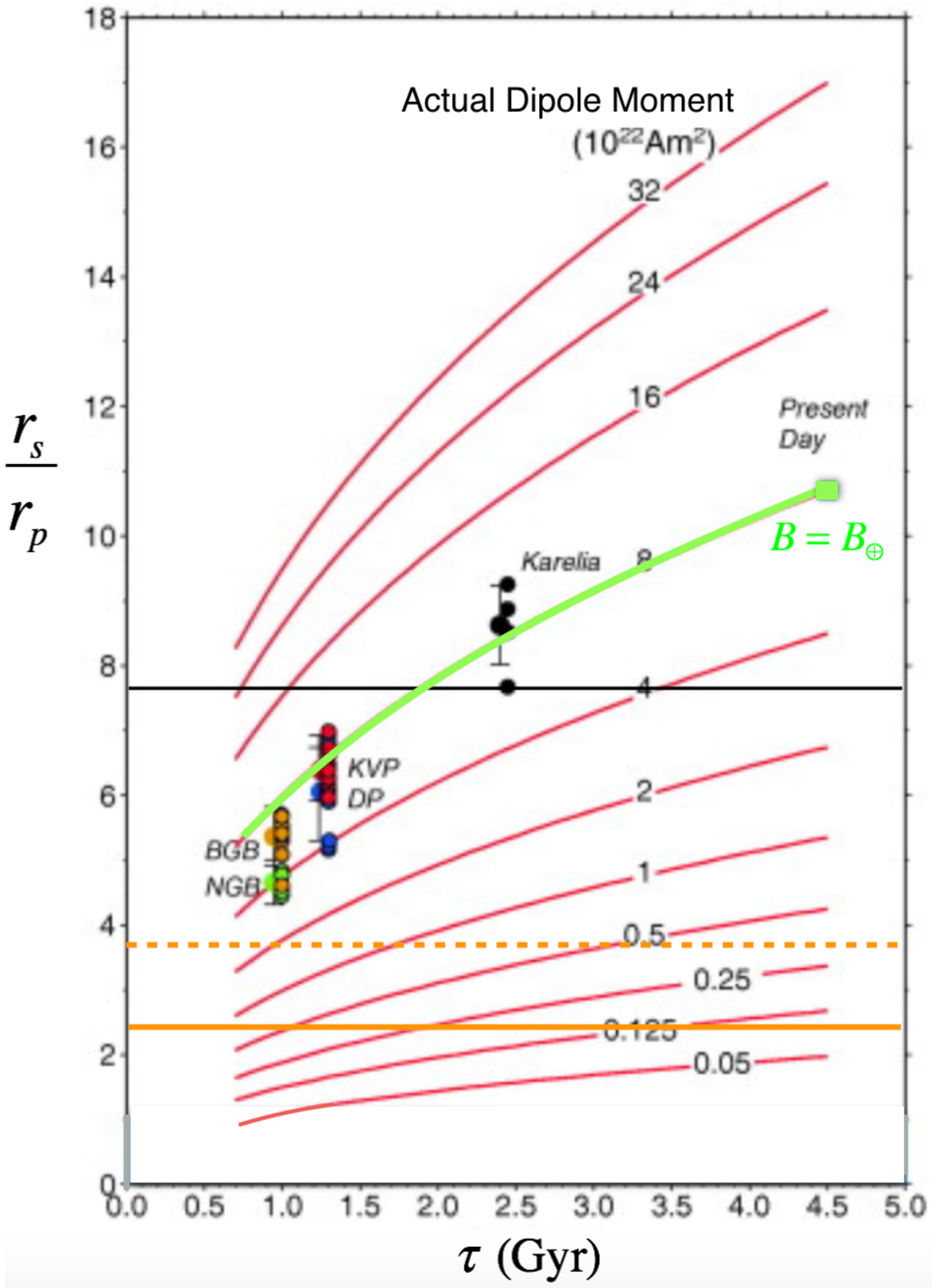}}
\caption{
Standoff distance in units of Earth-like planet radius vs.  age for different planet dipole moments  and   different wind evolution models:  top  \citep{Blackman2016}; 
bottom figure adapted from \citep{Tarduno2010}.
Green  lines of both panels  represent models  whose endpoint matches present day measurements.  Solar wind model curves in  bottom  panel result from combining mass-loss  to X-ray flux correlation with X-ray flux to age correlation (see text).
Magnetic dipole data sources for bottom panel are as follows: Karelia  \citep{Smirnov2003}, KVP, DP (Kaapvaal Pluton, Dalmien Pluton respectively) \citep{Tarduno2007}, and BGB, NGB (Barberton Greenstone Belt, Nondweni Greenstone Belt) \citep{Tarduno2010}. 
The geodynamo may have started shortly after the lunar forming event, when core chemical stratification was disrupted 
\citep[e.g.][]{Jacobson2017}.
 In both panels:    solid orange and solid black lines are   thresholds of  $r_s/r_p$  above which  mass   (Eq. (\ref{4.4.6}))
   and energy   (Eq. (\ref{en15a})),  respectively, per  solid angle captured toward the poles are higher  for a magnetized planet.  Dashed orange and dashed black lines are  the 
   corresponding thresholds
   for total wind mass (Eq. (\ref{4.4.5a})) and energy  (Eq. (\ref{en15}))
   respectively.
   Dashed black is absent on bottom panel due to the smaller  ordinate range.}
 \label{standoff}
\end{figure}




\subsection{Focusing toward the polar caps}

Magnetic polar caps subtend a solid angle  no larger than that of a hemisphere (2$\pi$)  and could be much less, as in the case of Earth.
When  captured particles funnel  toward  polar caps along the magnetic field rather than being uniformly distributed over a hemisphere,  the associated local
mass flux  (mass per unit time per area) and energy flux within the cap solid angle 
is  enhanced by  a  factor  $1/f_c$, where 
\beq
f_{c}= (1-cos \theta_c)\le 1
\label{fc}
\eeq
is the areal fraction of a hemisphere
subtended by a polar cap of colatitude  $\theta_c$.

If we restrict ourselves to  the case when the incoming wind flow is perpendicular to the planet's magnetic dipole axis, we can then estimate $\theta_c$ for a dipole from
the equation for the local  polar geometry of field lines, namely
 \citep{Russell2016+}
  \beq
r{ d\theta \over dr} = {B_\theta\over B_r}.
 \label{d1}
\eeq
For a vector dipole field in spherical coordinates of magnetic moment $\bf m$
\beq
{\displaystyle \mathbf {B} ({\mathbf {r} })={\frac {|\mathbf {m} |}{|\mathbf {r} |^{3}}}\left(2\cos \theta \, \mathbf {\hat {r}} +\sin \theta \,{\boldmath {\hat {\theta }}}\right)}, 
\eeq
from which the  ratio of the two field components is 
\beq
{B_\theta\over B_r}={tan\theta\over 2}.
 \label{d2}
\eeq
Using Eq. (\ref{d2}) in  Eq. (\ref{d1})  and integrating 
gives  
\beq
r=r_e sin^2\theta,
\label{theta}
\eeq
 where $r_e$ is the value of $r$ at which a particular field line crosses the equator $(\theta=\pi/2)$.
The  value $\theta=\theta_c$   is determined by setting $r_e=r_s$, the magnetopause radius,
and $r=r_p$ because  the field lines  anchored on the two polar circles $\theta = \theta_c$ at  $r=r_p$
represent the anchor loci of the  closed field lines 
just at the inner edge of the magnetopause.
Field lines anchored with  $\theta\le \theta_c$ extend into the wind-side of the magnetopause, along which wind plasma can only enter during reconnection. 

Using the aforementioned substitutions  in Eq. (\ref{theta}) along with Eq. (\ref{fc}) gives
\beq
f_c=1-\sqrt{1-r_p/r_s}.
\label{fc2}
\eeq
For a fixed wind mass flux, decreasing the magnetic field   increases  
$r_p/r_s$, thereby decreasing  $f_c$.  
Correspondingly,  for a  fixed planetary magnetic   field, $f_c$  decreases
with increasing wind ram pressure. This implies that
  $f_c$ would generally increase with stellar age (as wind power weakens) for steady planetary field strengths. 

We  define $q\equiv Q/f_c$ as the ratio  of mass  
capture rate per solid angle by a magnetized planet that includes polar cap channeling to
that in the absence of a magnetosphere altogether.
From  Eq. (\ref{4.4.5a})  we then obtain
\beq
q 
\simeq  {Q\over f_{c}}
\simeq 70.7\left({{\chi}\over 0.1}\right) \left({r_s\over 10r_p}\right)^{3}
\left[1+\left( 1-{r_p \over r_s}\right)^{1/2}\right],
\label{4.4.6}
\eeq
where the factor in square brackets ranges  between 1 and 2.
Note that 
 $q/Q=1/f_c$, which  increases with stellar youth.


Taken at face value,  Eqs. (\ref{4.4.5a}) and  (\ref{4.4.6})
 highlight that  the  mass   captured by a  magnetized Earth-like planet, when averaged over many  wind field reversals could  be  greater than without a magnetic field. 
{Our analysis here supersedes that of \cite{Tarduno2014} 
where we did not give explicit expressions for the reconnection speed and polar focusing
or emphasize that  
$Q$ and $q$ can exceed unity while the analogous energy
rate ratio can still be much less than unity. }

\subsection{Incoming energy flux with and without a magnetosphere}
  
The  ratio of energy input rates
 imparted to the atmosphere  for magnetized vs. unmagnetized planets
can be obtained by multiplying  Eq. (\ref{4.4.5a}) and Eq.  (\ref{4.4.6}) 
 by  
\beq
\xi={ \left({c_s^2\over \Gamma-1}+v_{in}^2\right)\over \left({c_s^2\over \Gamma-1}+v_{w}^2\right)}={ \left({c_s^2/v_w^2 \over \Gamma-1}+2{\chi}^2\right)\over \left({c_s^2/v_w^2\over \Gamma-1}+1\right)},
\label{enflux}
\eeq
where $c_s$
 is the incoming sound speed and $\Gamma$ is the adiabatic index.
The energy flux includes contributions from the enthalpy and the
bulk kinetic energy of the incoming material  with sound speed $c_s$.
Multiplying  Eq. (\ref{4.4.5a}) and Eq.  (\ref{4.4.6}) by  Eq. (\ref{enflux}) then gives
   the ratios of energy capture rates of the magnetized to unmagnetized planet for unfocused
   and polar focused cases  as 
 \beq
 Q_{en}\sim0.21\left({\xi\over 0.03}\right)\left({\chi\over 0.1}\right)
 \left({r_s/r_p\over 10}\right)^2,
 \label{en15}
 \eeq
 and
  \beq
  \begin{array}{l}
 q_{en}={Q_{en}\over f_c}\\\simeq 2.12 \left({\xi\over 0.03}\right) \left({\chi\over 0.1}\right)
 \left({r_s\over 10r_p}\right)^{3}
\left[1+\left( 1-{r_p \over r_s}\right)^{1/2}\right],
\end{array}
  \label{en15a}
 \eeq
respectively. 
The scaling  of $\xi =0.03$ arises if we use fiducial values for the  incoming wind of 
$c_s\simeq 3.9\times 10^6$cm/s (corresponding to $T\simeq 10^5$K),
$v_w=400{\rm km/s}$, and $\Gamma=5/3$.
 Eq. (\ref{en15})  shows that  the total energy collected by 
 a magnetized  planet scaled to present day Earth-sun-like parameters is  less than that of the magnetized planet. However  Eq. (\ref{en15a}) shows that the polar energy flux concentration
can be comparable, or even greater for the magnetized planet.
 
\section{Limits on Mass Loss Rates from Energy and Momentum Conservation}
\label{sec3}

The exact determination of the induced mass loss from an  incoming stellar wind  
depends upon detailed plasma and chemical properties that determine how the incoming energy and momentum  is converted to outgoing particles \citep{Sibeck1999,Lammer2013,Russell2016+}. We re-emphasize here that we are considering atmospheric mass loss estimates that involve the wind
 and not those arising only from  radiative forcing. The latter is needed to produce ions,
 but  atmospheric mass loss direct from only radiation  without the influence of the stellar wind  is    insufficient to explain the full history of mass loss
for Earth and Venus, and even Mars \citep{Dubinin2011+,Dong2018b+}.
 
 The extent to which an intrinsic planetary field can trap 
 outgoing particles  can also limit the otherwise induced  loss.
But regardless of  how the energy and momentum inputs are subsequently converted into outgoing particles, 
we can  compare estimates of  approximate maximum atmospheric mass loss rates  (``approximate" is explained below)  that the wind could induce  for  magnetized and unmagnetized cases. 
The distinction between  these two cases is essential when the exobase is below the magnetopause or ionopause.

 The wind energy flux times the capture area for inflow speed $v_{in}$ is given by
\beq
\rho_{in} Av_{in}\left({c_s^2\over \Gamma-1}+v_{in}^2\right)={\dot M}_{in} \left({c_s^2\over \Gamma-1}+v_{in}^2\right),
\eeq
and the wind momentum flux  times the capture area is 
\beq
\rho_{in}A \left( v_{in}^2+{c_s^2\over \Gamma}\right)={\dot M}_{in}v_{in} \left(1+{c_s^2\over \Gamma v_{in}^2}\right).
\eeq

   
The maximum  outflow induced mass loss rate   corresponds to the case in which all of  the incoming momentum or energy gets converted into outflow with  outflowing ions moving just  at the escape speed $v_{esc}$ 
along open field lines. For   thermal, isotropically  distributed ions,   the relevant
initial  ion speed would be the most probable particle speed at the exobase $v_p$.  If 
$v_p$  were already even as high as a fraction  $\lambda=1/4$ of the escape speed $v_{esc}$ before any wind influence, then for a  Maxwellian  distribution $f(v)=({-mv^2\over2\pi kT})^{3/2}e^{-mv^2\over2 kT}$, 
only a fraction 
${\int_{ v_{p}/\lambda}^{\infty}  f(v) v^2 dv\over  \int_{0}^{\infty}  f(v) v^2 dv} \sim 7.4 \times 10^{-9}$
would be above the escape speed, and the fraction  drops exponentially with decreasing $\beta$.
When this fraction times the atmospheric density at the exobase is less than that of the stellar wind,
 the number of particles above the escape speed would be negligible.

But even particles  with  speeds   as high as $v_{esc}/4$, have only 1/16th of the escape energy.  Therefore, assuming that particles start from zero energy does not lead
to a significant difference in estimating  a mass loss rate.
Moreover, for cases when the particle gyro-radii are initially much  smaller than the planet scale,  then only motion along open field lines  leads to escape. Since the escape path  is locally   along 1 direction,  no more than 1/3 of the particles   escape when the most probable speed reaches the escape speed for an isotropic distribution.  To ensure that all particles   escape if they were initially at $1/16$ the escape energy then requires  $3-1/16\sim 3$ times the escape energy to be imparted from the incoming wind.
 
If the ratio of the inflow speed to the escape speed exceeds unity, then energy conservation gives a larger outflow mass loss rate limit than momentum conservation.
 Otherwise, momentum conservation determines the larger mass loss rate limit.
For a steady state,   the corresponding energy and momentum rate balance equations
for inflow and outflow are given respectively by 
\beq
{\dot M}_{in} \left({c_s^2\over \Gamma-1}+v_{in}^2\right)
= 
  {\dot M}_{out} v_{esc}^2,
 \label{enint}
\eeq
and 
\beq
{\dot M}_{in}v_{in} \left(1+{c_s^2\over \Gamma v_{in}^2}\right)
=
{\dot M}_{out} v_{esc}.
 \label{momint}
\eeq
By rearranging  each of Eq. (\ref{enint}) and  (\ref{momint})
 to obtain an expression for 
 ${\dot M}_{out}$, we can combine
 the results to express  the maximum of the two upper limit estimates for the outflow mass loss. If we also allow for the polar focusing fraction $f_c$, the combined result is 
  \beq
  \begin{array}{l}
{ {\dot M}_{out,max}\over {{\dot M}_{in}}}\\ 
={1\over f_c} {v_{in}\over v_{esc}}\left[{{v_{in}\over v_{esc}}\left(1+{c_s^2/v_{in}^2\over \Gamma-1}\right)}, \left(1+{c_s^2/v_{in}^2\over \Gamma }\right)\right],
\end{array}
\label{outboth}
 \eeq
 where the first term in brackets is the result from energy conservation and the second term is from momentum conservation. The former dominates when $v_{in}> v_{esc}$, and the latter when 
 $v_{in}< v_{esc}$. If all the energy and momentum instead went into  particles 
 moving at some speed $v_{ave}> v_{esc}$, then $v_{ave}$ would replace $v_{esc}$ in 
 Eq.   (\ref{outboth}) and the mass outflow rate estimate would be correspondingly reduced.
 


%

 \section{Implications for Atmosphere Protection}
 \label{sec4}
We now apply the formalism of the previous 
sections to the time evolution  of magnetospheric protection for the 
Earth-sun system, and to the  separate question of present day oxygen ion 
loss rates from Earth, Mars, and Venus.
 
\subsection{Influence of  stellar activity variation on  protection by an Earth-like magnetosphere}


The blue curves of the top panel and the red curves of the bottom panel of Fig. \ref{standoff} show the time evolution of $r_s/r_p$ for different values of the planetary magnetic field (assumed to be constant for each curve) for an Earth-like planet at 1AU from a sun-like star. The top panel uses the minimalist, but holistic theoretical model of \citep{Blackman2016} to compute the standoff distance from the  wind ram pressure. The latter is  derived from the time evolution of magnetic field, rotation, mass loss, and X-ray luminosity predicted from the physics of the model.
Although the correlation between  X-ray luminosity and stellar mass loss has been questioned \citep{Cohen2011}, this query is based on short-term observations of the Sun that may not apply to the billion year time scales of interest. The bottom panel in Fig. \ref{standoff} shows curves from \cite{Tarduno2010} which utilize the empirical mass loss values for solar-like stars from  \citep{Wood2006} based on H I Lyman-$\alpha$ data. Those values are observed to correlate with X-ray flux, which in turn was used to empirically determine stellar ages. 
 The magnetopause standoff distances are truncated at 700 my because the few available young stellar analogs hint at lower rates of mass loss \citep{Wood2014}, breaking the correlation.
(The saturation of X-ray flux with youth also  makes  
age determination more difficult for younger stars \citep{Mamajek2008,Wright2011}.)

The green curve in the top panel of  Fig. \ref{standoff}  corresponds to the case that matches  present day field measurements, as also  labeled on the bottom panel.
The curves of both panels show that  $r_s/r_p$ decreases with youth 
because of the stronger ram pressure of the solar wind, while the planetary field remains nearly constant.
The horizontal lines in the two panels of Fig \ref{standoff} correspond to the specific thresholds calculated in Section {\ref{sec2}.
Namely,  the orange solid lines  show the threshold of $r_s/r_p$ 
above which  $q\ge 1$ from Eq. ({\ref{4.4.6}) (assuming the lower limit of $1$ for the  square bracketed quantity).
 The orange dashed lines show  the  threshold above which $Q\ge 1$ from  Eq. (\ref{4.4.5a}).
These are, respectively, the thresholds above which the polar and total atmospheric mass collection rates for the magnetized planet exceed those for the unmagnetized planet with a compact atmosphere.
The  solid and dashed black curves, which correspond
to Eqs  (\ref{en15a}) with the  square bracketed quantity equal  to $1$, 
and (\ref{en15}) respectively, 
 indicate the threshold ratio of $r_s/r_p$
above which the polar focused and total energy  capture rates by the magnetized planet exceed that for  an unmagnetized
planet.  Because energy conservation rather than momentum conservation determines the  larger of the two values on the right of  Eq. (\ref{outboth}) for Earth-sun-like
systems, these curves also indicate the corresponding thresholds above which the magnetized planet  has  higher upper limits on mass loss from stellar-wind atmosphere erosion as compared to the unmagnetized planet.

Given the fiducial parameter values used to make the horizontal lines, 
both panels of Fig. \ref{standoff}  reinforce the same  implications: 
for most of  an Earth-like planet's history,  more total mass and  polar mass flux can be captured by the magnetized planet than for  the correspondingly unmagnetized planet, even though  less total energy is captured by the magnetized planet over its past history.  As shown,  this circumstance will also persist for the the long term future. 
Since the maximum mass outflow rate for an Earth-sun-like system is determined by energy rather than momentum in Eqn. (\ref{outboth}),  this  further implies that the upper limit on  the mass loss rate would be reduced by a magnetosphere throughout the  Earth-like planet's past and future.

 If we include polar focusing using $f_c$ from Eq. (\ref{fc}) however, the solid black horizontal lines of Fig \ref{standoff} exemplify that  the polar  focused  energy flux (and thus the resulting upper limit on  atmospheric mass outflow 
from Eq. (\ref{outboth}))
 can be {\it higher} in the recent past and future  for the magnetized Earth compared to an  unmagnetized Earth.  
Here the two panels differ somewhat in where the line crosses the  
standoff curve that best matches present day  but as alluded to above, neither the data nor the models are   accurate enough for this time scale to be constrained to better than a factor of 2.

Irrespective of the cross-over time discussed above, 
Fig. 2 shows, for an Earth-like magnetized planet, that the initial outgoing mass loss due to interaction with the wind 
 (and ignoring return flow) can be larger with a magnetosphere than without under certain conditions because the magnetosphere
can capture and focus more incoming wind energy than would otherwise have access to
the atmosphere.
 How close the actual mass loss corresponds to  the maximum 
 limit implied by Eq. (\ref{outboth})  depends on rate of ion return and on how incoming energy and momentum interact with specific planetary atmospheres. 
 We touch on the latter in the next section.




\subsection{Partial  explanation of  Oxygen ion loss from Mars, Venus, and Earth}


The question of why $O^+, O^{+}_2$    loss rates on Earth, Mars and Venus  are apparently within a few orders of magnitude of each other is of interest. This  has led  some to question the traditionally assumed protective role of an internally generated magnetosphere \citep{Strangeway2010,Wei2012,Strangeway2017+} since Earth has one but Mars and Venus do not.
The detailed processes and local conditions of different atmospheres are time dependent, and the measured rates   depend on solar activity phase and spacecraft  location, so there is a range of more than an order of magnitude even  for a given planet.  
The complexity of the underlying systems  
ultimately require  combining the chemistry and microphysics to obtain detailed models. 
The diversity of numerical simulations of this sort include
 M-star stellar winds interacting with habitable magnetized, and 
  Venus-like planets, to obtain  standoff distances and  lower limits on mass loss
 \citep{Cohen2014,Cohen2015};
studies of the  Trappist exoplanet system \citep{Dong2018a+}; and simulations of 
the time evolution of Mars' oxygen loss  \citep{Dong2018b+} that presently agrees with   recent MAVEN data \citep{Jakosky2015+}.


{ We  will use  our formalism to obtain  approximate upper limits on oxygen ion escape rates for Earth, Mars, and Venus-like planets  from  interaction with the solar wind, using energy and momentum  
 conservation separately. We will  compare the results for  internally magnetized (Earth-like) and non-internally magnetized (Mars and Venus-like) cases.\bf }
 \cite{Zendejas2010+},  presented a different minimalist model  of ion outflow rates, using an imposed efficiency parameter $\alpha$ to characterize how much inflow energy is converted to outflow.    A value  $\alpha=1$ would make their results also upper limits, but they do not address the case of a magnetized planet.

We will see that the  atmospheric mass loss estimates   that result from use of momentum conservation in 
our approach are in fact similar for Earth, Venus, and Mars to the level of an order of magnitude. Our limits that result from use of energy conservation are similar for the three planets to within two orders of magnitude. 


\subsubsection{Upper Limits for Earth}
 To obtain an atmospheric mass loss limit ${\dot M}_{out} \le {\dot M}_{out\oplus}$
for an Earth-like magnetized planet,  we consider the fluxes as integrated over solid angle and thus  avoid $f_p$ here.
We then use Eq. (\ref{vrecvin}), 
 Eq. (\ref{4.4.2}), and energy conservation in Eq. (\ref{outboth})  to obtain
\beq
\begin{array}{l}
 {\dot M}_{out\oplus}=7\times 10^4\left({\dot M_w\over 1.3 \times 10^{12}{\rm g/s}}\right)\left({v_w\over 4\times 10^7{\rm cm/s}}\right)^2\left({\chi\over 0.1}\right)^3 \\
 \left({r_s \over10  r_\oplus}\right)^2\left({R\over 1{\rm AU}}\right)^{-2}
\left({r_{ex}\over 1.2 r_\oplus}\right)\left({M_p\over M_e}\right)^{-1}\left({{{c_s^2/[2\chi^2v_w^2(\Gamma-1)]} +1}\over 1.64}\right){\rm g\over  s}.
\end{array}
\label{mdotplanet}
\eeq
If instead we use  momentum conservation in Eq. (\ref{outboth}),  we  obtain
\beq
\begin{array}{l}
 {\dot M}_{out\oplus}=
1.5\times 10^4\left({\dot M_w\over 1.3 \times 10^{12}{\rm g/s}}\right)\left({v_w\over 4\times 10^7{\rm cm/s}}\right)\left({\chi\over 0.1}\right)^2 \\
 \left({r_s \over10  r_\oplus}\right)^2 \left({R\over 1{\rm AU}}\right)^{-2}
\left({r_{ex}\over 1.2 r_\oplus}\right)^{1\over 2}\left({M_p\over M_e}\right)^{-{1\over 2}}\left({{{c_s^2/(2\chi^2v_w^2\Gamma)]} +1}\over 1.26}\right){\rm g\over s},
\end{array}
\label{mdotplanetmom}
\eeq
where  $M_p$ is the planet mass, 
$r_{ex}$ is the  exobase radius, which,  along with  $r_s$ are
scaled to Earth's radius $r_\oplus$.
For Earth,   $r_{ex}\sim 1.2 r_\oplus$ and $r_s\sim 10r_\oplus$, ${\chi}=0.1$.
From Eq. (\ref{outboth}), the corresponding  particle loss rates limits from Eq. (\ref{mdotplanet}) and Eq. (\ref{mdotplanetmom}) are  respectively 
\label{26}\beq
dn/dt \le  {{\dot M}_{out\oplus} \over \mu m_p}\simeq 
2.62 \times 10^{27}  \left({\mu \over 16}\right)^{-1} {\rm s}^{-1},
\label{27}\eeq
and
\beq
dn/dt \le  {{\dot M}_{out\oplus} \over \mu m_p}\simeq 
5.6 \times 10^{26}  \left({\mu \over 16}\right)^{-1} {\rm s}^{-1}.
\label{28}\eeq
This can be compared with 
$7.2 \times 10^{25}$/s of $O^+$ observed with speeds above 
escape leaving the poles, and averaged over  a solar cycle, of which 90\% seems to be   recaptured \citep{Seki2001}. 


\subsubsection{Upper Limits for Venus and Mars}


For Mars and Venus,  photoionization and dissociative recombination convert  neutral molecules  into $(O^+$ or $O_2^+)$  and these collisionless ions can then be accelerated  by the motional electromotive force  \citep{Lundin+1992,Dubinin2011+,Russell2016+}  near the ionopause. 
 These ions then drift perpendicular to both the incoming wind  velocity and   magnetic field.  Since the incoming flow has no initial net momentum perpendicular to the flow, lateral acceleration of oxygen ions is  accompanied by a momentum conserving flow in the opposite direction. Electrons can play this role when ion-gyro-radii are small, 
but in a multi-ion-species plasma with  gyro-radii comparable or   larger than system scales,  proton streaming in the  opposite direction to the oxygen atoms  is also important 
\citep{Dubinin2011+}.  

As  oxygen atoms drift away from the subsolar point, their  Larmor orbits  take them into the  wind plasma where they are picked up by the oncoming wind,  accelerated and  carried past the planet.  For Venus, the ion production mechanisms likely do not  directly accelerate ions to escape speeds so the wind-atmosphere interaction is essential 
but for Mars the initial energy of the ions from dissociative
recombination might produce some escape ions directly.
Photochemical dissociation may  dominate 
presently,  but ion-pickup could dominate for  ancient Mars \citep{Dong2018b+}.  As  discussed in section 3, even if the ions initially have speeds as high as 1/4 the escape speed before acquiring wind energy, the  requirements to escape would  be nearly the same as if particles started from zero energy.
 


Focusing first on  Venus, we note that  standoff  is 
determined by the ionopause particle pressure, but the induced magnetic field piles up just outside this region to a value in near equipartition with the
internal thermal pressure \citep{Russell2016+}.
The induced magnetopause boundary  is of order $50$ km or so above the  ionopause,
and the latter is  only $\sim 250$ km above the surface. 
Since the ionopause, magnetopause, and exobase are all only within a few $100$ km above the surface,  they are relatively small distances above the planet radius $r_p$.  However, the influence of the wind on atmospheric mass loss requires coupling of the wind and ionosphere.
Once the ions are picked up by the wind, this coupling is determined by 
finite  gyro-radii--an estimate of which will be needed to determine
 $r_T$  in the cross section for interaction between wind and atmosphere.

To quantify the atmospheric mass loss, we  first apply Eq. (\ref{4.4.1}) for ${\dot M}_{in}$  in combination with Eq. (\ref{outboth}) and $v_{in}=v_w$ 
to obtain $ {\dot M}_{out} \le {\dot M}_{out\venus}$. Using energy conservation  in Eq. (\ref{outboth}), we obtain
\beq
\begin{array}{l}
 {\dot M}_{out\venus}=
{1.5\times10^6}\left({\dot M_w\over 1.3 \times 10^{12}{\rm g/s}}\right)\left({v_w\over 4\times 10^7{\rm cm/s}}\right)^2
\left({R\over 0.72{\rm AU}}\right)^{-2}\\
\left({\pi r_T^2\over 8\times 10^{18}{\rm cm^2} }\right)
\left({r_{ex}\over 0.95 r_\oplus}\right)\left({M_p\over 0.81M_e}\right)^{-1}\left({{{c_s^2/[v_w^2(\Gamma-1)]} +1}\over 1.01}\right){\rm g/s}.
\end{array}
\label{mdotplanet2}
\eeq
If instead we use  momentum conservation in Eq. (\ref{outboth}), we obtain 
\beq
\begin{array}{l}
 {\dot M}_{out\venus}= 
3.86\times 10^4
\left({\dot M_w\over 1.3 \times 10^{12}{\rm g/s}}\right)\left({v_w\over 4\times 10^7{\rm cm/s}}\right)
\left({R\over 0.72{\rm AU}}\right)^{-2}\\
\left({\pi r_T^2\over 8\times 10^{18}{\rm cm^2} }\right)
\left({r_{ex}\over 0.95 r_\oplus}\right)^{1/2}\left({M_p\over 0.81M_e}\right)^{-1/2}\left({{{c_s^2/(v_w^2 \Gamma)} +1}\over 1.00}\right){\rm g/s},
\end{array}
\label{mdotplanet2mom}
\eeq
where in both Eq. (\ref{mdotplanet2}) and (\ref{mdotplanet2mom}) we have
used $r_T$ as defined in Eq. (\ref{rT}) with $r_g$ computed for 
$O^+$ ions  at the solar wind speed  using a Parker spiral field  \citep{Parker1958}
 estimate for the  uncompressed  wind magnetic field  $B(R)= B_\odot (R_\odot^2/ R) (\Omega_\odot/v_w)$ at Venus' distance from the sun.  This provides a maximum estimate  for $r_g$ since a  larger field and smaller thermal speed lowers $r_g$.  The result is
\beq
r_g=  9.9\times 10^8 {\rm cm}\left({\mu\over 16}\right)\left({v_w\over 4\times 10^7{\rm cm/s}}\right)\left({B\over 7\times 10^{-5} {\rm G}}\right)^{-1}.
\label{31}
\eeq
For $\mu=16$, the  corresponding ion loss rates are then
\beq
dn/dt \le  {{\dot M}_{out\venus} \over \mu  m_p}\simeq
  5.58 \times 10^{28}
   {\rm s}^{-1}
\label{32}\eeq
and 
\beq
dn/dt \le  {{\dot M}_{out\venus} \over \mu  m_p}\simeq 
1.44\times 10^{27}
  {\rm s}^{-1}.
\label{33}
\eeq
This  latter limit is an order of magnitude  larger than the highest values reported in \citep{Knudsen1992,Dubinin2011+} for ions above the escape energy.

For Mars, aside from its localized magnetic features, much of the surface has too
little magnetic or ionospheric pressure to steadily abate the incoming wind ram pressure \citep{Russell2016+}.
We again use Eq. (\ref{mdotplanet2})
but now with the Martian parameters of $M_p =0.11M_e$; $r_p\sim r_s=r_{ex}=0.53 r_\oplus$, 
and $R=1.5AU$ in Eqn. (\ref{outboth}).
For  $r_T$ we  use  Eq. (\ref{rT}) but 
 employing Mars' radius for  $r_p$ 
and   $r_g$  is now the gyro-radius for $O_2^+$ ions moving at  the solar wind speed in  the uncompressed solar wind field at Mars, namely, 
$
r_g=  4.2\times 10^9 \left({\mu\over 32}\right)\left({v_w\over 4\times 10^7{\rm cm/s}}\right)\left({B\over 3.3\times 10^{-5} {\rm G}}\right)^{-1}{\rm cm}$.

The resulting  mass loss limit from  energy conservation from Eq. (\ref{outboth}) is then
${\dot M}_{out\mars}=1.3\times 10^{5}$g/s and 
for $\mu=32$, the corresponding ion loss rate is then
\beq
dn/dt \le  {{\dot M}_{out\mars} \over \mu  m_p}\simeq 
2.3\times 10^{29}
 {\rm s}^{-1}.
\label{34}
\eeq
If instead we use  momentum conservation in Eq. (\ref{outboth}), then we obtain
${\dot M}_{out\mars}=1.6\times 10^{3}$g/s and the corresponding ion loss rate  limit of
\beq
dn/dt \le  {{\dot M}_{out\mars} \over \mu  m_p}\simeq
  2.9 \times 10^{27}
  {\rm s}^{-1}.
\label{35}
\eeq
These can be compared with values up to
$2\times 10^{25}/{\rm s}$ at solar max reported in
\cite{Dubinin2011+}.  

Comparing the results from Eqs. (\ref{27}) \& (\ref{28});  (\ref{32}) \& (\ref{33}); and 
and (\ref{34}) \& (\ref{35}) we see that (i)
the  wind-induced mass loss  limit estimates from energy conservation span 2 orders of magnitude  with ${\dot M}_{out\mars} > {\dot M}_{out\venus}> {\dot M}_{out\oplus}$,  and all exceed the respective limits from momentum conservation for each of the three cases;  (ii)  the mass loss limits from momentum  conservation  satisfy the same inequality ordering but   all are within an order of magnitude of each other;   (iii)   the momentum limits are all appropriately  larger than the largest measured values.   It is a worthy pursuit to identify the basic principles that dictate how efficiently the incoming energy and momentum are redistributed among particles and on what this depends.  Momentum limits may be more pertinent than energy limits for  unmagnetized planets as the flow has more direct access to the atmosphere. In addition, we might speculate that the maximum efficiency with which momentum is converted to oxygen ions given the aforementioned comparisons of predicted versus measured values is of the order of 10\% for Earth and Venus and 1\% for Mars.  The latter is perhaps consistent with the possibility emerging from at least one global model, namely  that photodissociation may dominate Mars' present ion loss, but
that of ancient Mars  ($\sim 4$ Ga) was two orders of magnitude larger, and dominated by ion-pickup \citep{Dong2018b+}.  That said, the relative role of photodissociation
vs. ion-pickup is still an area of active study.
However,  given the current limitations on observed loss rates and incoming wind flux,
 these pursuits are left  for future study.


In short, our formalism  may help to explain aspects of the present-day loss rates of  terrestrial planets with atmospheres.  We reiterate that this is separate from the  specific question of the role of magnetic fields in sustaining or inhibiting habitability. 
The  present state of Venus and Mars 
reflects a longterm atmospheric evolution with  a profound loss of the principal component defining habitability--water.  Present day Venus is particularly inhabitable.
As such, the  present-day atmospheric erosion rates  do not provide  direct  measures of  depletion of a specifically habitable atmospheres but
still offer   testbeds for  basic principles of 
atmospheric mass loss given different sources of external stellar forcing.
Those principles are still important to identify even if they do not comprise
a  complete set of ingredients that determine habitability. }

\section{Conclusions}

We have considered how a planetary magnetosphere influences the upper limit on stellar wind induced mass loss  by focusing on   how the magnetic field affects  incoming flow  of 
mass and energy into the atmosphere. The approximate maximum outflow rates are estimated separately for the constraint of energy conservation and momentum conservation assuming that  (i) all of the incoming energy or momentum goes into  outgoing particles just at the escape speed; and (ii) that the atmospheres have initially thermal energy distributions  with most probable speeds initially less than 1/4 the escape speed. Our  formalism  can be used to obtain  upper limit estimates for a given planetary magnetic field, planet size, mass, orbital radius and stellar wind parameters. 
For the parameter regimes considered, momentum conservation provides a lower upper limit than energy conservation.

The  formalism may be helpful toward a better understanding of the role and extent that magnetic fields play in protecting atmospheres of  both solar system and extrasolar planets.
In particular, we find that the competing  
effects of decreased solar wind flux interacting with the atmosphere versus increased solar wind collection area predict that (i) for most of Earth's history, more total solar wind mass may have been captured with potential to interact with the atmosphere than would have been the case without a geodynamo but that (ii) the geomagnetic field still had a net protective from the solar wind  because less wind energy was input as compared to would be the case of an unmagnetized Earth.  
However, when polar focusing is taken into account,  the future  polar energy flux collected by Earth's magnetosphere could exceed that which would be collected without a field.  The latter implies that the future protective effect of the geomagnetic field will depend primarily on its ability to recapture \citep{Seki2001} otherwise ablated plasma.  
We also showed how the competing effects of decreased solar wind flux interacting with the atmosphere versus increased solar wind collection area
might contribute to explaining the explain present-day oxygen ion ejection rates from Earth, Mars, and Venus.

Since the retention of water, and therefore oxygen, is a  key ingredient for Earth-like habitability,  our results highlight that the influence of planetary magnetic fields on the survival of  habitable atmospheres must ultimately depend on the relative interplay  between  incoming wind flux reduction,  increase wind capture area, and outgoing atmosphere re-capture.

\section*{Acknowledgments} 
 { We thank  J. Carroll-Nellenback and J. E. Owen for useful related discussions and the referee
for thoughtful and  useful comments.   
  EB acknowledges  support from National Science Foundation (NSF) and 
  Hubble Space Telescope (HST) grants 
 NSF-AST-1109285 and HST-AR-13916.002, respectively, and from the Simons Foundation,  the Institute for Advanced Study (Princeton) and the Kavli Institute for Theoretical Physics of UCSB with support from NSF grant PHY-1125915.
 JAT acknowledges  support from grants NSF-EAR-1656348
 and NSF-EAR-1520681.}

\bibliographystyle{mn2e}
\bibliography{standoff}

\end{document}